\documentclass{PoS}

\title{Towards closure testing of parton determinations}

\ShortTitle{Towards closure testing of parton determinations}

\author{\speaker{Nathan P. Hartland}\\
        University of Edinburgh\\
        E-mail: \email{n.p.hartland@ed.ac.uk}}

\author{Christopher S. Deans\\
        University of Edinburgh\\
        E-mail: \email{c.s.deans@sms.ed.ac.uk}}

\abstract{The method of closure testing for analysing the effectiveness of a PDF fitting procedure is discussed. In order to pass a closure test, a fitting methodology must be able to reproduce a known generating function in a fit to an ideal pseudo-dataset generated by that PDF up to the level of experimental uncertainty in the data. Here we present an initial study of the closure property of the NNPDF fitting methodology. An idealised pseudo-dataset is generated by a set of toy PDFs that differ substantially from previous NNPDF determinations. In a fit to this pseudodata, the NNPDF methodology is shown to be able to reproduce well the toy PDFs used as a generating function. }

\FullConference{XXI International Workshop on Deep-Inelastic Scattering and Related Subjects -DIS2013,\\
		22-26 April 2013\\
		Marseille, France}

\begin{document}
Given the importance of parton distribution functions (PDFs) to the physics program of the LHC, a rigorous understanding of the fitting procedures used in the determination of PDFs and an estimation of their methodological reliability is vital. The NNPDF collaboration has developed a methodology which provides an unbiased approach to PDF determination and has produced several PDF sets with this methodology. The latest such set, NNPDF2.3~\cite{Ball:2012cx} is the first PDF determination to include LHC data.


To obtain a reliable determination of PDFs, and in particular to provide a robust account of their errors, a sufficiently flexible parametrisation at the fitting scale is required. Indeed, recent results
from the MSTW group demonstrated that an extended parametrisation improved the description of $W$ lepton asymmetry data from the LHC~\cite{Martin:2012da}. An integral part of the NNPDF methodology is the use of artificial neural networks to provide the initial scale fitting basis, providing a great deal of flexibility in the functional form. This removes a possible source of bias in the fit due to the parametrization being overly restrictive. 

The properties of the NNPDF methodology have been previously studied in detail in a number of contexts. The fit results have been shown to be largely independent of the particular architecture chosen for the neural networks, and shown to be stable under a significant enlargement of the number of parameters in the fit~\cite{Ball:2011eq}. The Monte Carlo approach to PDF uncertainties used in NNPDF fits has also been demonstrated to behave in a statistically robust fashion under the addition of new data by use of a Bayesian reweighting analysis~\cite{Ball:2010gb}. The contribution of functional form flexibility to PDF uncertainties has also been quantified~\cite{Ball:2011eq}.

Here we shall discuss a method of directly evaluating the efficacy of a fitting methodology and the flexibility of a PDF parametrisation by studying a fit to an idealised dataset. This provides a clean environment where the capabilities of a fitting procedure may be analysed. Firstly a set of pseudodata points corresponding to real experimental data used in the NNPDF 2.3 fit are generated from the predictions of an existing PDF set. The experimental data values used in the fit are then replaced by these theoretical predictions leading to a set of data central values that are internally consistent and produced by a known generating function. Experimental noise is then added to the pseudodata according to the covariance of the corresponding experimental data. The final step is to perform a fit to the pseudodata and compare the PDFs obtained to the generating function. This tests the closure property of a fitting methodology, in that the resulting fit should be able to reproduce the generating function to within the variation expected from the experimental uncertainties.

The primary benefit of performing such a closure fit is that we are able to test the flexibility of a fitting procedure in an environment where the underlying theory is known. The PDF set used to generate the pseudodata provides a `correct answer' which can be compared to the resulting closure test fit. Additionally, the procedure for generating the pseudodata guarantees that the data will be completely self-consistent and with fluctuations which match those expected from the experimental covariance matrix. The closure test therefore provides an environment for testing the methodology which is free of data inconsistencies and under- or over-estimated systematics. Other tests can also be performed where inconsistencies are added to the pseudodata, by shifting particular datasets, rescaling or removing systematics, which test the impact of such anomalies in PDF fits. Closure tests have previously been used by the MSTW group to assess the behaviour of PDF errors under the introduction of such artificial data inconsistencies\cite{Watt:2012tq}.

While this is a test that may be performed with NNPDF fits in the future, here we present preliminary work which looks at the ability of the NNPDF methodology to reproduce a known PDF used as a generating function. As an initial closure test, a full NNPDF fit has been performed to a set of pseudodata generated by the Les Houches benchmark toy PDFs \cite{Giele:2002hx}. These benchmark PDFs differ considerably from the results of previous NNPDF determinations therefore providing a significant test of the flexibility of the fitting procedure and parametrisation. The pseudodata was generated consistently at NLO by the same calculations that are used in the fit itself.

\begin{figure}[tbp]
\begin{center}
 \includegraphics[scale=0.7]{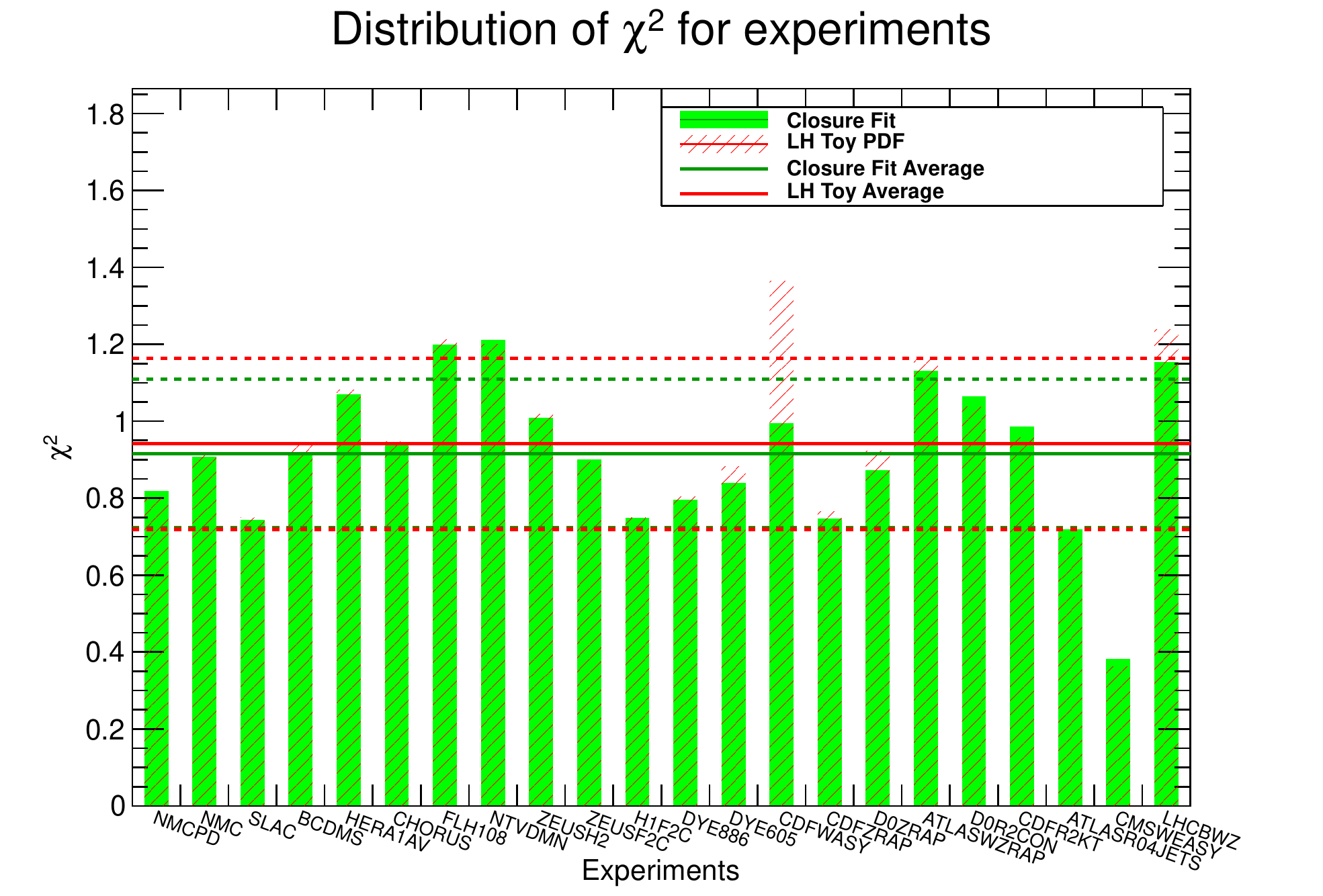}
\caption{ $\chi^2$ values to pseudodata for the closure test fit and the generating function (Les Houches toy PDFs) for each of the different datasets involved in the fit. The average $\chi^2$ and one standard deviation are also shown. It can be seen that with a few possible exceptions the closure test fit is able to reproduce the $\chi^2$ of the generating function well.}
\label{chi2pseudo}
\end{center}
\end{figure}

In terms of fit quality, the closure test fit is able to reproduce well the LH toy PDF's description of the data. To pseudodata, the generating function has a global $\chi^2$ of $0.95$, and the closure fit obtains a $\chi^2$ of $0.94$. In Figure \ref{chi2pseudo} we see the comparison of $\chi^2$ values to pseudodata, broken down dataset by dataset in the fit. Note that the $\chi^2$ of the pseudodata even to the PDF set which generated it is not uniformly one. This is due to the random fluctuations introduced during the data generation. It is clear however, that with the possible exception of the CDF W asymmetry data the closure test fit is able to reproduce the toy PDF well at the level of fit quality to the pseudo-dataset. 

\begin{figure}[htbp]
\begin{center}
 \includegraphics[scale=0.375]{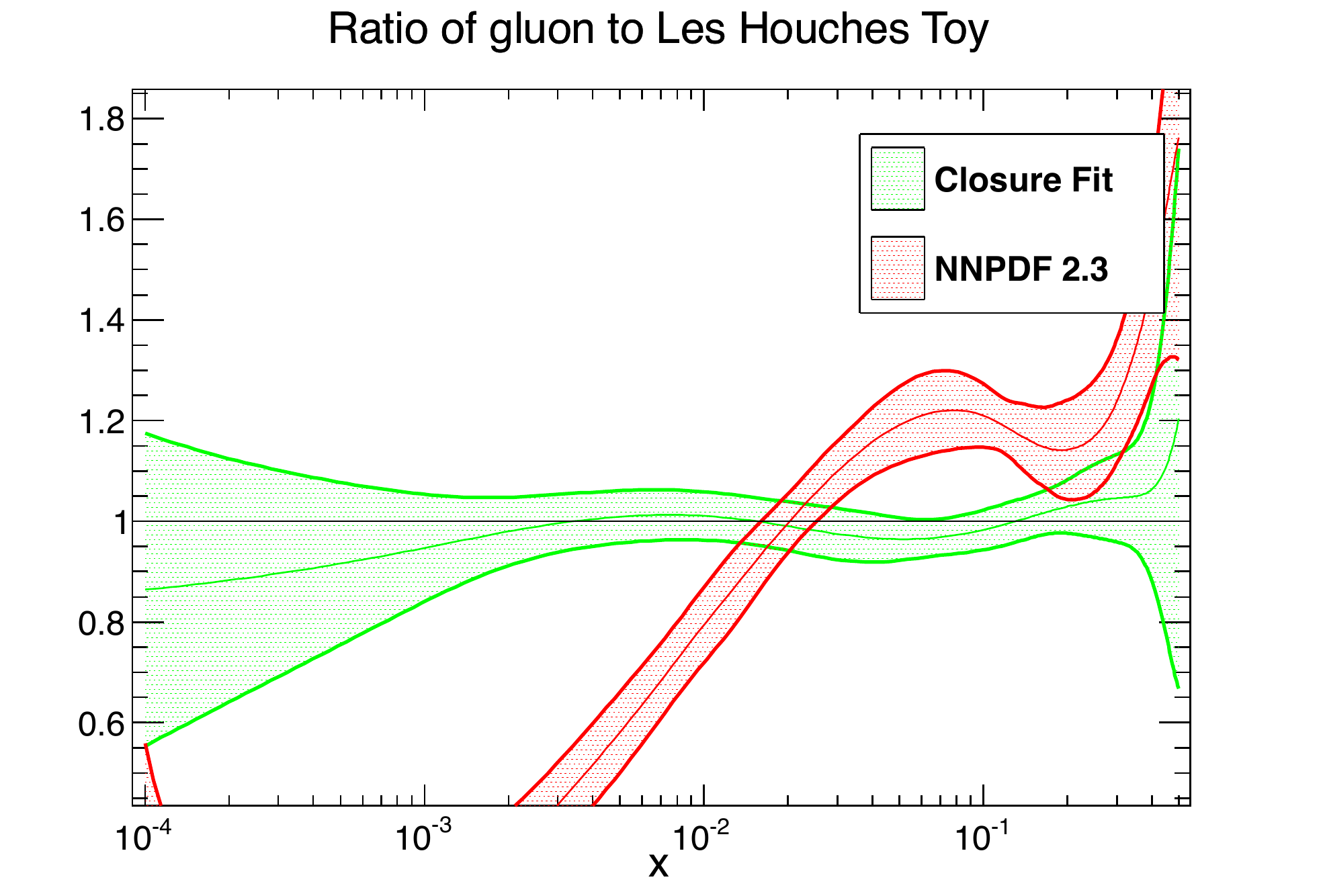}  \includegraphics[scale=0.375]{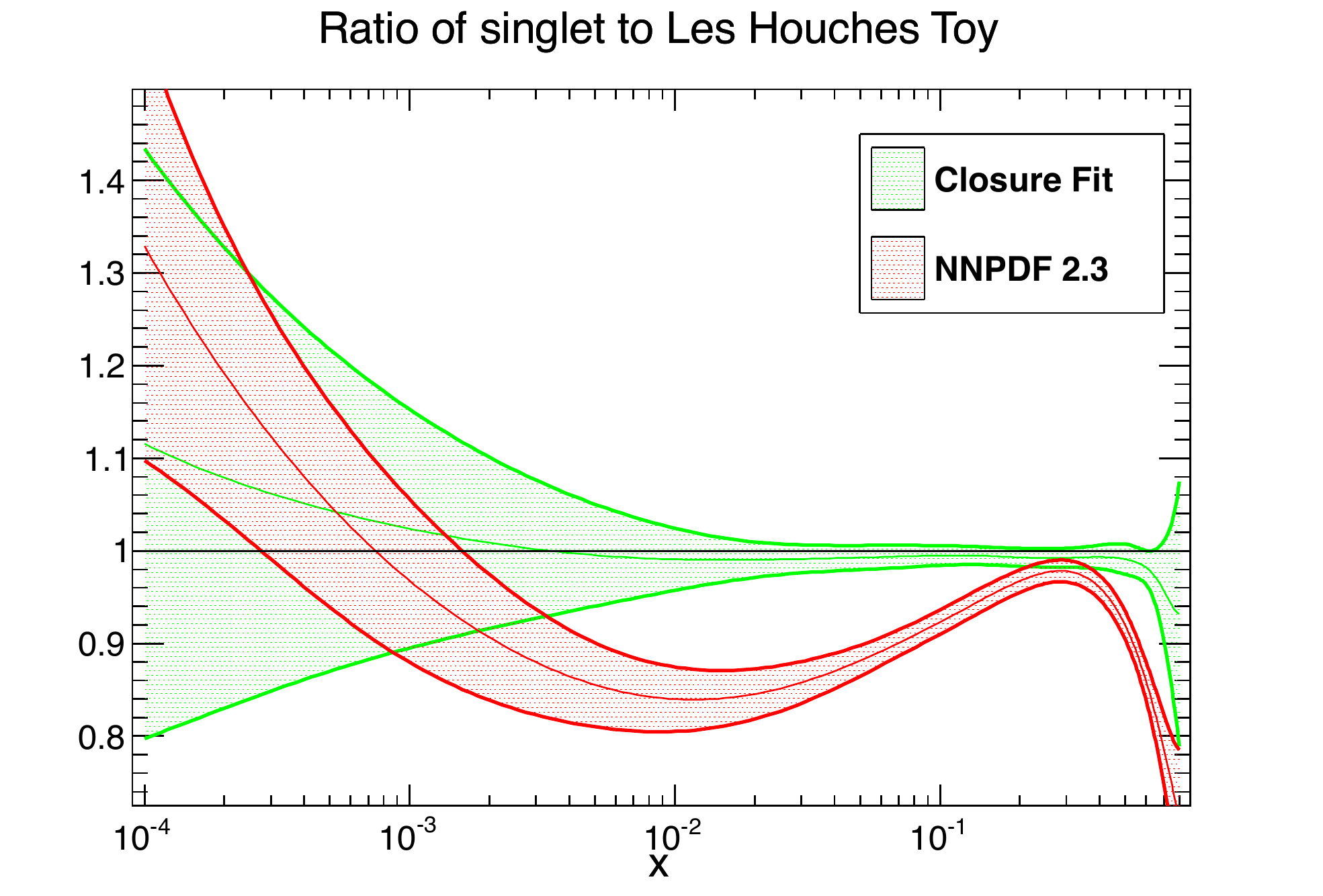}
\caption{Ratio of the gluon and singlet PDFs from the closure test to the equivalent Les Houches toy PDFs used to generate the closure test psuedodata. The central value and one sigma bands are shown. In both PDFs the closure test reproduces the generating PDFs to within one sigma. The ratio for the NNPDF2.3 PDFs are also shown. }
\label{toyfits1}
\end{center}
\end{figure}

\begin{figure}[htbp]
\begin{center}
  \includegraphics[scale=0.375]{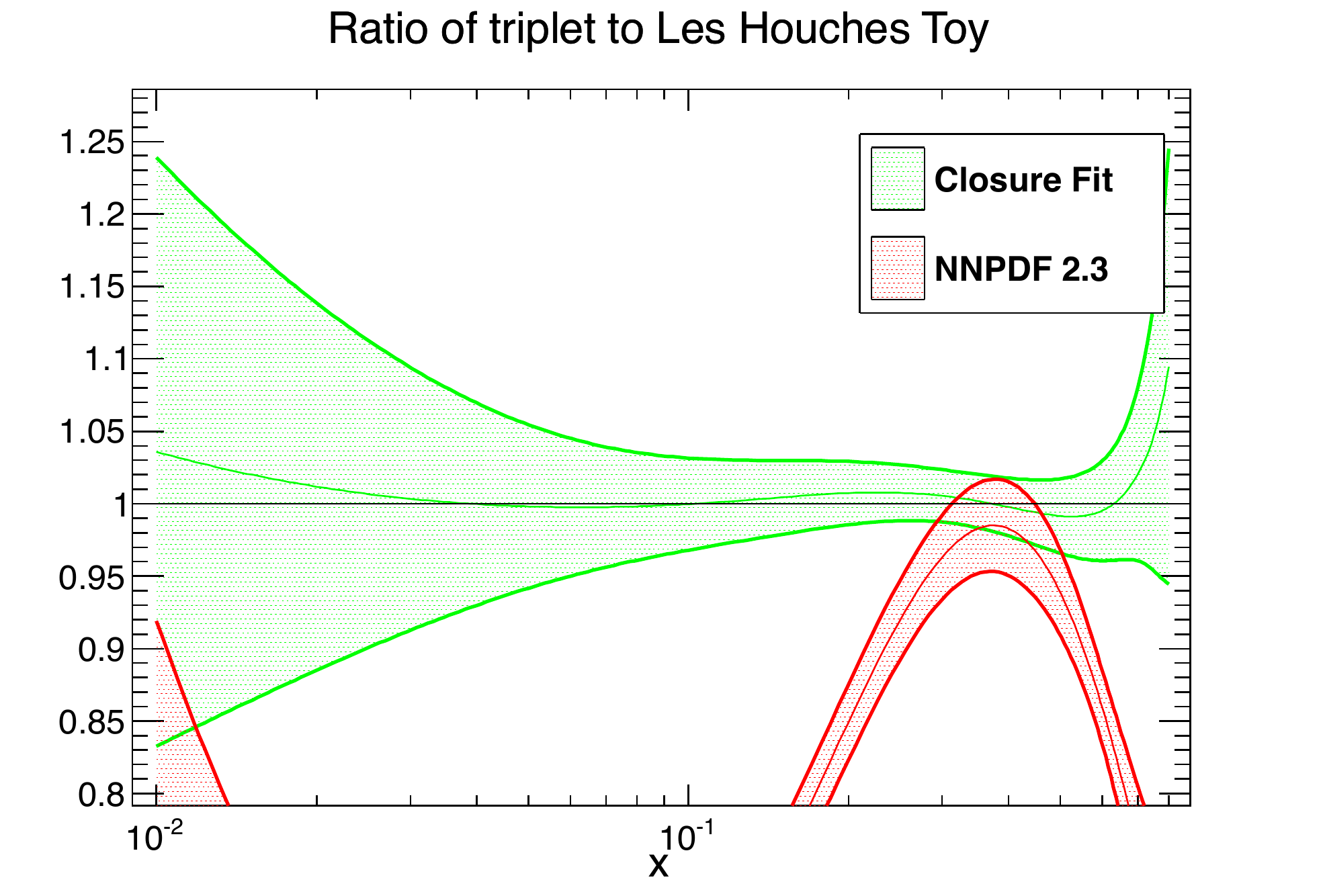}  \includegraphics[scale=0.375]{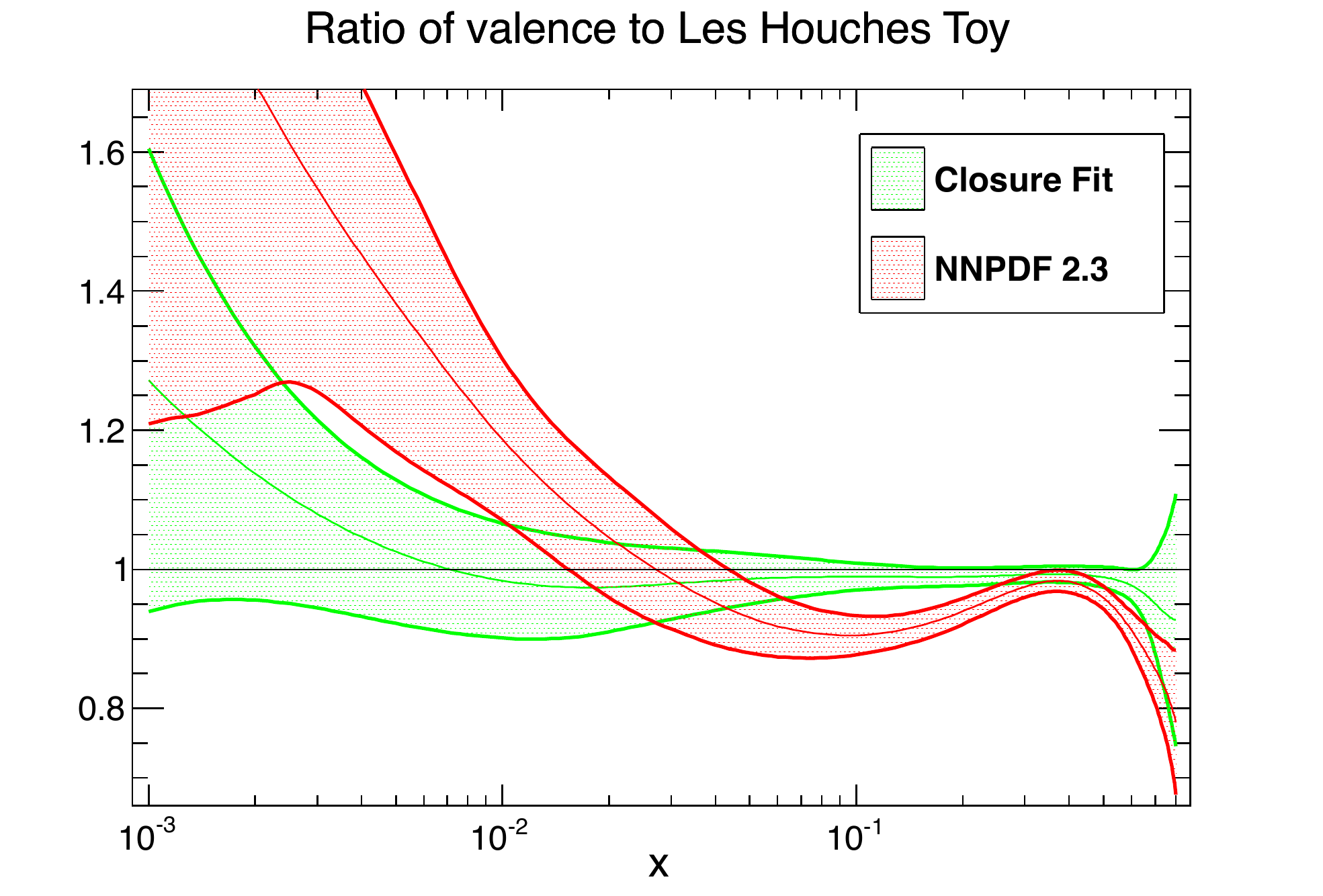}
\caption{Same as figure 3 but for the triplet and valence PDFs. }
\label{toyfits2}
\end{center}
\end{figure}

Having assessed the ability of the closure test to reproduce it's generating function at the level of fit quality, it is instructive to examine agreement at the level of the PDFs themselves.
Figures \ref{toyfits1} and \ref{toyfits2} show ratios of the closure test PDFs (in green) and the NNPDF 2.3 PDFs (in red) to the Les Houches toy PDFs. In order to pass the closure test criterion, the fit to pseudodata should contain the generating function within it's uncertainty band. We can see from the figures that the closure test reproduces the generating set consistently to within one standard deviation. The closure test is able to reproduce the generating function even in regions where the toy PDFs depart by many standard deviations from the result of the NNPDF2.3 fit to experimental data. This demonstrates the flexibility available in the NNPDF procedure.

\begin{figure}[htbp]
\begin{center}
\includegraphics[scale=0.5]{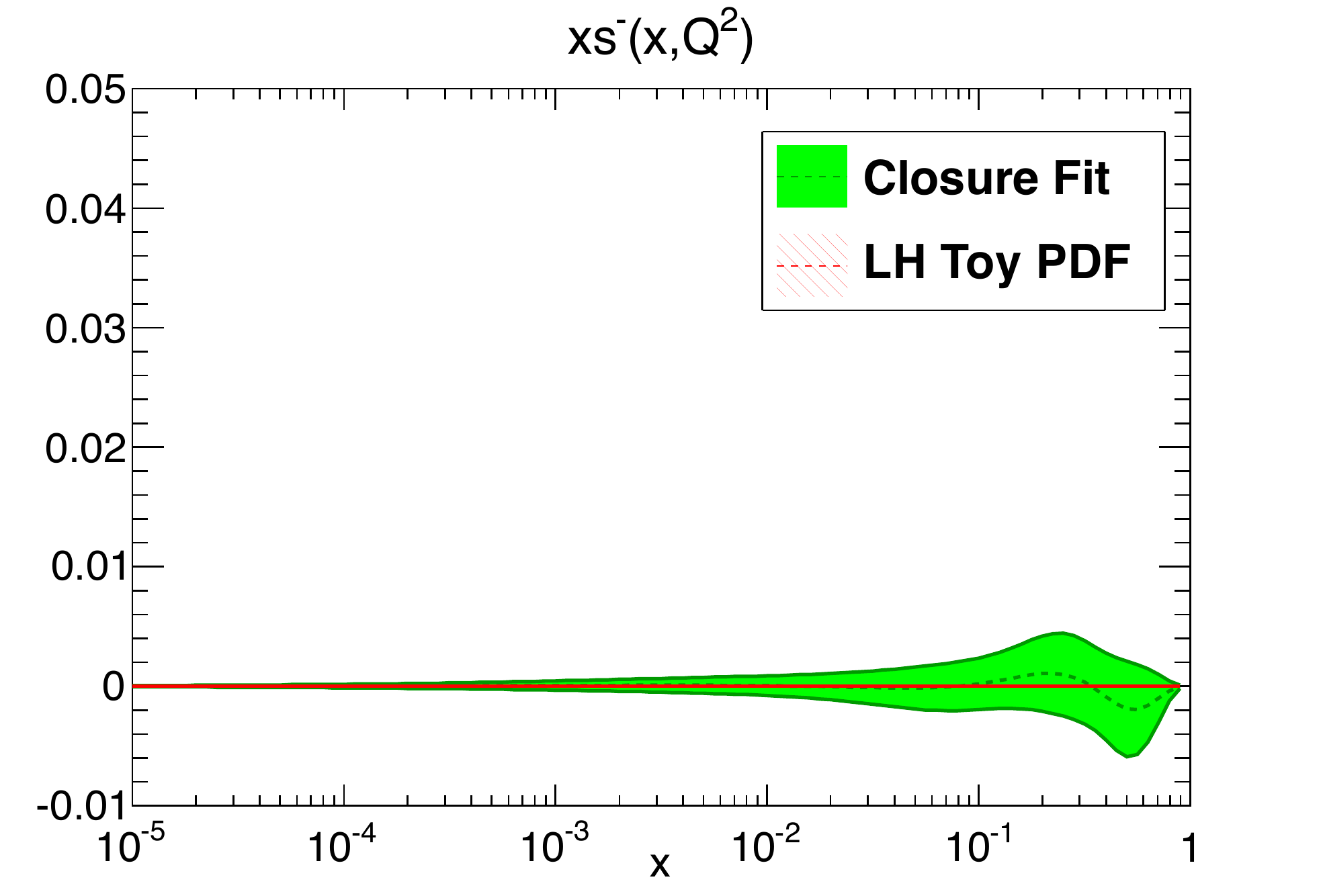}
\caption{Strange valence PDFs for the closure test fit and the Les Houches toy PDFs. The dotted line is the central value while the bands show the one sigma contour. The closure test fit correctly determines a zero strange valence within one sigma.}
\label{toyfits3}
\end{center}
\end{figure}

Figure \ref{toyfits3} shows that the strange valence distribution $s^{-} = s - \bar{s}$, assumed to be zero in the Toy PDF Set but parametrized independently in the NNPDF methodology is reproduced well. This demonstrates that the fitting procedure is able to correctly discern a zero strange valence distribution if the data suggests it, reaffirming that the structure found for this distribution in standard NNPDF fits is driven by the experimental data rather than parametrization inflexibility.

\clearpage

These results suggest that the NNPDF fitting methodology can successfully reproduce a known underlying theory which has a very different form from the standard NNPDF fits. The parametrisation used appears to be flexible enough to cope with the Les Houches Toy PDFs used as a generating function. The requirement that a fitting procedure should successfully pass a closure test is a powerful check on the reliability of the methodology. A large amount of new data relevant to the determination of PDFs is now available from LHC collaborations, the closure test technique will be applied to further refine the NNPDF methodology to ensure a reliable determination of parton densities under the addition of these new datasets in the next major NNPDF release.

\end{document}